# Energy landscape in NiCoCr-based middle-entropy alloys

Nikolai A. Zarkevich,[1] Timothy M. Smith,[2] John W. Lawson[1]

[1] Intelligent Systems Division, NASA Ames Research Center, Moffett Field, CA 94035, USA
[2] NASA Glenn Research Center, 21000 Brook Park Rd., Cleveland, OH 44135, USA

**Abstract**:
NiCoCr middle-entropy alloy is known for its exceptional strength at both low and elevated operating temperatures. Mechanical properties of NiCoCr-based alloys are affected by certain features of the energy landscape, such as the energy difference between the hcp and fcc phases (which is known to correlate with the stacking fault energy in the fcc phase) and curvature of the energy surface. We compute formation energies in the Ni-Co-Cr ternary and related quaternary systems and investigate dependences of the relative energies on composition. Such computed composition-structure-property relations can be useful for tuning composition and designing next-generation alloys with improved strength.

## 1. Introduction:

Among multi-principal element alloys (MPEAs),[1] the NiCoCr middle-entropy alloy (MEA) has remarkable strength[2] and ductility,[3] especially at low and high operating temperatures $T$.[4] Due to their competitive mechanical properties and corrosion resistance,[5] MPEAs find applications in engines and heat exchangers in aerospace,[6,7] automotive,[8] nuclear,[9,10] and energy[11] industries. Such alloys are also suitable for additive manufacturing.[12-15] There are attempts to further improve strength[16] by adjusting the composition and adding precipitates, such as oxides.[17] Properties depend on microstructure, which is affected by composition and heat treatment.[18] Although novel alloys with improved creep and strength are synthesized and investigated,[19] theoretical understanding of the dependence of their properties on composition is still lacking. Here we provide results of density functional theory (DFT) combined with multiple scattering theory (MST) and predict the dependence of formation and relative energies on composition.

The enhanced strength-ductility balance in NiCoCr and related MPEAs is believed to result from peculiarities in the microstructure,[19] such as locally variable stacking fault energies,[20] which correlate with the energy difference between the face centered cubic (fcc) and hexagonal close-packed (hcp) structures. Locally, the hcp phase looks like an intrinsic stacking fault in the fcc structure, while a double hexagonal close-packed (dhcp) structure represents an extrinsic stacking fault. Similarly, $D0_{19}$ and $D0_{24}$ structures locally look like intrinsic and extrinsic stacking faults in the $L1_2$ phase,[21] relevant for precipitation in Ni superalloys.[22] Interestingly, there is no ordering phase transition in $Ni_3(Ti_{0.9168}Al_{0.0832})_1$ alloy, which has a frustrated ground state and an inherent atomic disorder on the Ti/Al sublattice.[21] Although not every disordered alloy is a compositional glass,[21] atomic disorder can be frozen at low temperatures. Diffusion depends on composition and temperature $T$ and typically becomes negligible at low $T$. In a doped alloy, defects interact with the solute and can either attract or repel each chemical species. Hence, the local chemical composition at and near defects can differ from that in the bulk. The energy surface near a defect can be dimpled.[23] Defects can be point (0-dimensional), linear (1D), planar (2D), or bulk (3D). Examples of defects include vacancies (0D), dislocations[20] (1D), surfaces, grain boundaries, twins, stacking faults (2D), and precipitates (3D). The mobility of defects affects the mechanical properties of an alloy.[24] The behavior of defects, and atomic diffusion near defects, are governed by relative energies and temperature. Computation of the energy-composition dependencies in disordered alloys from the first principles is challenging.[25] Here we address this challenge.

We combine several theoretical and computational methods for dealing with disordered alloys.[26] We apply them to homogeneous solid solutions and construct the composition-structure-property relations, which can be useful for materials design.[27]

This article is organized as follows. Methods are described in section 2. Results and discussion are in section 3. Summary (section 4) is followed by the Appendix.

## 2. Methods:

Density functional theory (DFT) is used to compute the electronic structure and energies of ordered and homogeneously disordered structures.[26] The ground state (GS) structure of Ni is face-centered cubic (fcc); the GS of Cr is body-centered cubic (bcc); and the GS of Co is hexagonal close-packed (hcp), see Fig. 1. We consider 3 phases: fcc (A1), bcc (A2), and hcp (A3). Both fcc and hcp phases are close-packed and differ by the periodic stacking of atomic layers, which is [ABC] along the cubic [111] direction in fcc and [AB] along the hexagonal [0001] direction in hcp.

DFT is combined with multiple scattering theory (MST)[28] to address atomic disorder.[26] Homogeneous atomic disorder is considered within the coherent potential approximation (CPA),[29] combined with the Korringa-Kohn-Rostocker (KKR) method.[30,31] Electronic-structure calculations are performed in primitive unit cells using the all-electron KKR-CPA code[32] with the periodic boundary corrections. The hcp structure is considered with the ideal $(c/a)=(8/3)^{1/2}$ within the KKR method. We checked that the computed ground state structures of elemental Ni, Cr, and Co are fcc, bcc, and hcp, respectively, see Fig. 1.

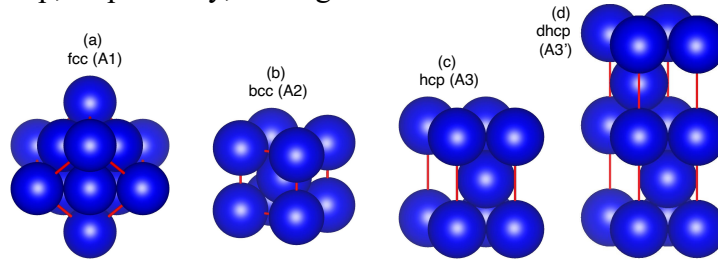

Figure 1. The fcc (A1, cF4, Cu, 225), bcc (A2, cI2, W, 229), hcp (A3, hP2, Mg, 194), and dhcp (A3', hP4, α-La, 194) structures (with Strukturbericht designation, Pearson symbol, prototype, and space group number). The fcc, hcp, and dhcp structures are close-packed and differ by a periodic stacking of atomic layers: [ABC] along the cubic [111] direction in fcc and [AB] or [ABAC] along the hexagonal [0001] direction in hcp or dhcp structures, respectively.

At each composition, 5 points with 1.5% step in the lattice constant were used to fit the Birch-Murnaghan[33,34] equation of state (EoS), from which the equilibrium volume $V_0$ and energy $E_0$ were determined and plotted versus composition $x$. The $E_0(x)$ data was interpolated using either polynomials or cubic splines. Triangulation and cartesian interpolation were used to create ternary contours.

The VASP full-potential code[35,36] with the projector augmented-wave method[37] was used to compute energies of elemental solids. Assuming that the systematic error is linear versus composition in each phase within the KKR method, the full-potential corrections were applied to the KKR results. The PBEsol[38] exchange-correlation functional was used in both VASP and KKR codes.

### 3. Results and Discussion:

The equiatomic NiCoCr alloy is known to have the fcc structure at room temperature (RT) in experiment. We find that its ground state is hcp, while the hcp and fcc homogeneously disordered NiCoCr solid solutions have similar energies at 0K.

The computed total and projected on atomic spheres electronic density of states (DOS) in fcc (A1) NiCoCr is shown in Fig. 2. At the Fermi energy $E_F$, the total DOS and the atom-projected electronic DOS of Ni and Co decreases with energy $E$, while that of Cr increases with $E$ at $E_F$ in all 3 phases. The total spin-projected DOS is similar in fcc (A1) and hcp (A3) phases, and these phases have similar energies, while bcc (A2) phase is higher in energy at this composition, see Fig. 3.

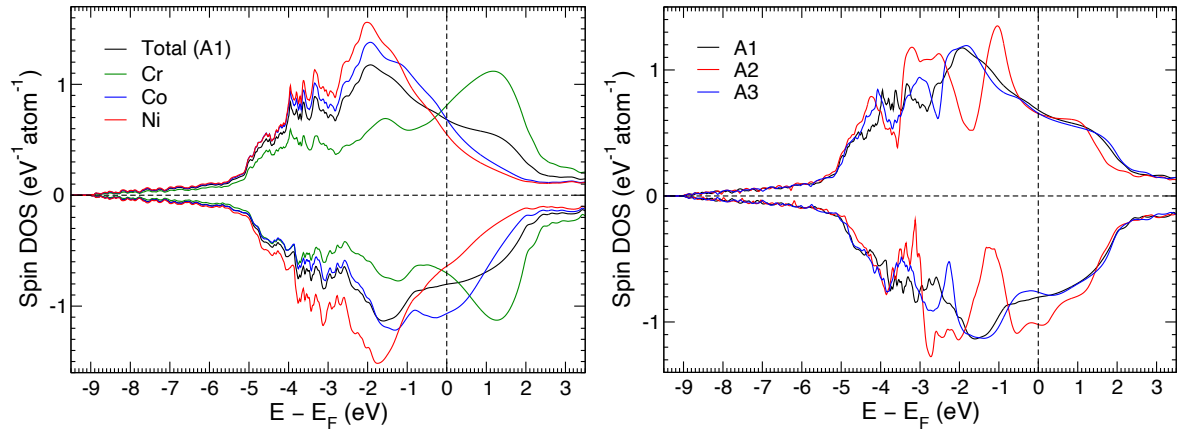

Figure 2. Electronic density of states in equiatomic NiCoCr: total and projected on the atomic spheres in fcc phase (a), and total in fcc (A1), bcc (A2), and hcp (A3) phases (b).

The computed formation energy of NiCoCr is positive, see Fig. 4. Thus, development of microstructure and atomic short-range order[39] must be energetically favorable in experimental samples.[40] Relative energies are shown in Fig. 5, while formation energies and equilibrium volumes (per atom) along selected cross sections of the compositional space are plotted in Figures 6 and 7 for the Ni-Co-Cr ternary system and Figures 8 and 9 for the quaternary $(CrCoNi)_{1-x}X_x$ alloys with X=(Ti, Zr, Hf; V, Nb, Ta; Mo, W; Re; Al, Cu, Fe).

Interestingly, at $T=0$ K the computed energy of the hcp NiCoCr homogeneously disordered solid solution is slightly lower than that of fcc, while a small fcc-hcp energy difference of 2.67 meV/atom is comparable to the DFT error (of the order of ~1 meV/atom). The higher-symmetry cubic fcc phase has a higher lattice entropy than the hexagonal hcp phase. Entropy is expected to play a role in stabilizing the fcc phase at and above the room $T$. The computed formation energy $E_0(fcc)=+143.6$ meV/atom is positive for NiCoCr. For a homogeneously disordered ternary equiatomic alloy, the mixing entropy is $S_{mix}/k_B = \ln(3) = 1.0986$ per atom, where $k_B$=0.08617333 meV/K is the Boltzmann constant. At room $T\approx25°C$ (298 K), $TS_{mix}$=28 meV, small compared to $E_0$=+143.6 meV/atom for the fcc NiCoCr. Thus, the free energy ($E_0 - TS_{mix}$)≈+116 meV/atom is positive at RT, and NiCoCr is expected to be chemically inhomogeneous, with energetically favorable deviations from homogeneity. Both energy and mixing entropy are lowered by an atomic ordering, such as a short-range order (SRO) or a long-range order (LRO).

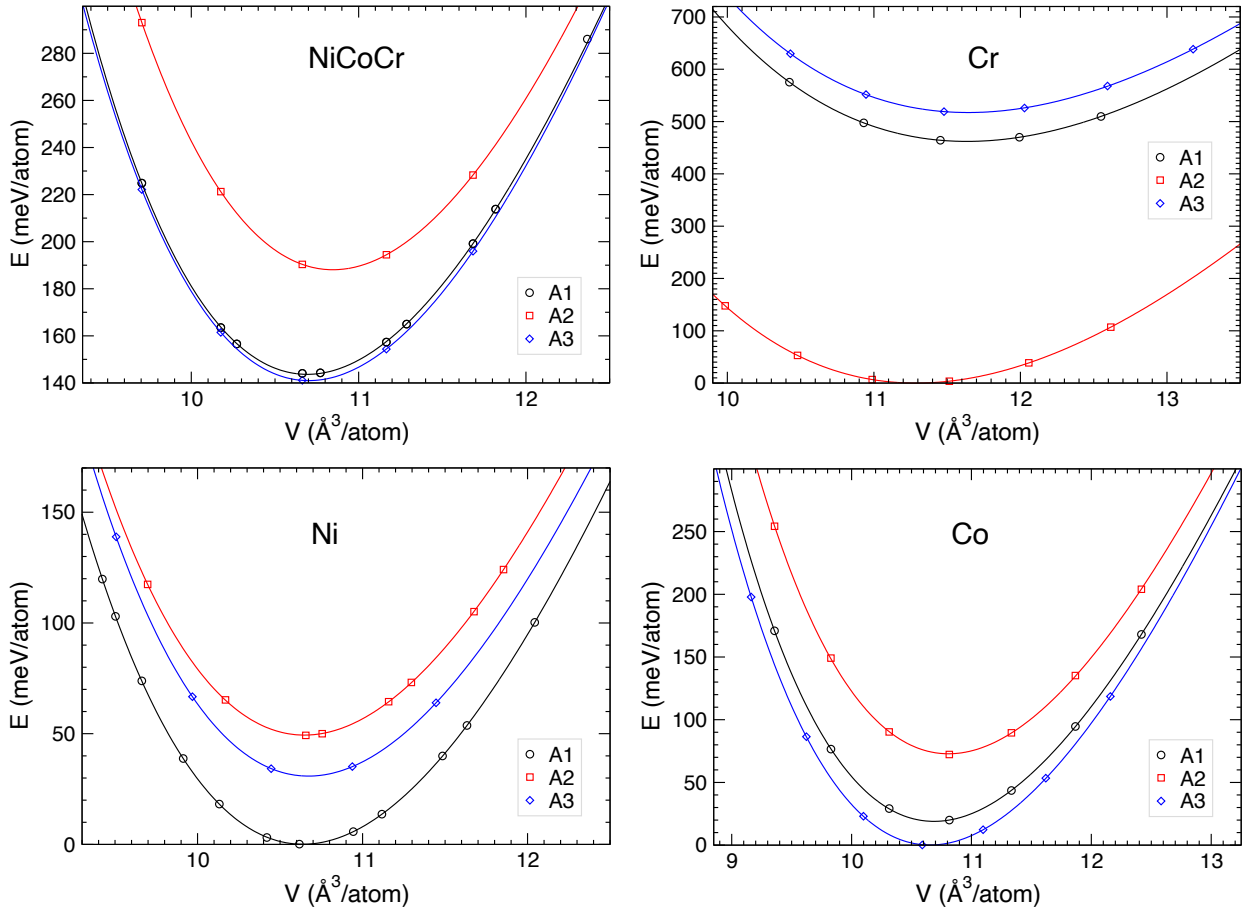

Figure 3. Computed formation energy versus volume in fcc (A1), bcc (A2), and hcp (A3) phases for the equiatomic NiCoCr and the elemental Ni, Co, and Cr. DFT points are fitted by the Birch-Murnaghan EoS (lines).

The computed equations of state (EoS) in fcc (A1), bcc (A2), and hcp (A3) phases for the equiatomic NiCoCr middle-entropy alloy are compared with those for the elemental solids in Figure 3. One can see that the ground states of Ni, Cr, and Co are A1, A2, and A3, respectively, while A1 and A3 close-packed structures have very similar energies in NiCoCr at $T$=0K. Because A3 (hcp) structure locally looks like a stacking fault in A1 (fcc), the energy difference between hcp and fcc phases correlates with the stacking fault energy, which is low in NiCoCr; this result agrees with previous computations[41] and experimental measurements.[42,43]

For the equiatomic NiCoCr, the computed at $T = 0$ K equilibrium fcc lattice constant is $a_0$= 3.497 Å; it is 2% smaller than the experimental[44] $a$ = 3.564 Å, measured by the XRD at room $T$, see Table A1. The computed at $T = 0$ K NiCoCr bulk modulus is $B_0$=232 GPa, see Table A2. The computed bulk moduli of the elemental Ni, Co, and Cr are 218, 227, and 260 GPa, while the measured $B$ values are 220 (Ni), 206 (Co), and 280 GPa (Cr), respectively. Interestingly, $B(NiCoCr) \approx [B(Ni)+B(Co)+B(Cr)]/3$ = 235 GPa. However, the computed at 0K value of $B_0$=232 GPa for the homogeneously disordered NiCoCr solid solution differs from the experimentally assessed value for a sample with microstructure and atomic ordering. In experiment,[24] a sample was obtained by arc-melting, re-melted five times, drop-casting the melt into a copper mold, and growing single crystals from the polycrystalline as-cast ingots using a floating-zone directional solidification method. For the single-crystal NiCoCr sample,[24] the experimental values of the elastic constants are $c_{11}$=249 GPa, $c_{12}$=156 GPa, $c_{44}$=142 GPa, the bulk modulus is $B=(c_{11}+2c_{12})/3$=187 GPa, and the Young's modulus is $E$=234 GPa at room temperature. The experimental[24] assessment of Debye temperature was 490 K in NiCoCr.

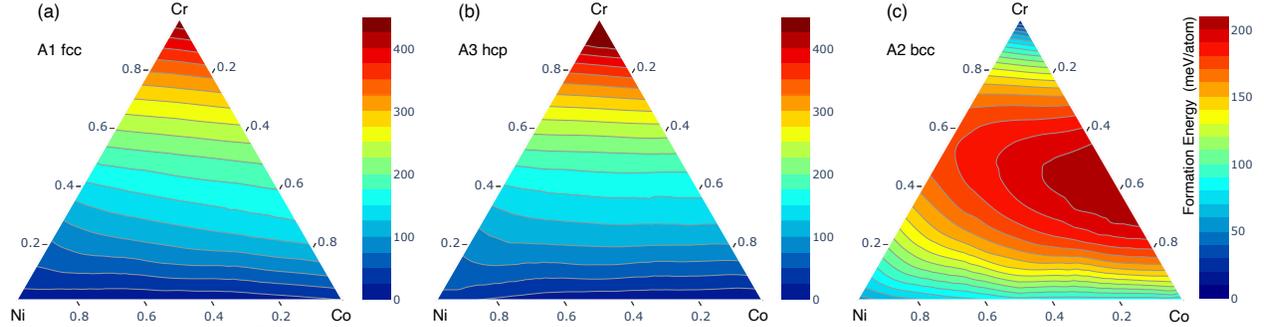

Figure 4. Computed formation energy [meV/atom] at equilibrium volume versus composition (atomic fraction) in Ni-Co-Cr ternary alloys in fcc, hcp (same scale), and bcc (different energy scale) phases.

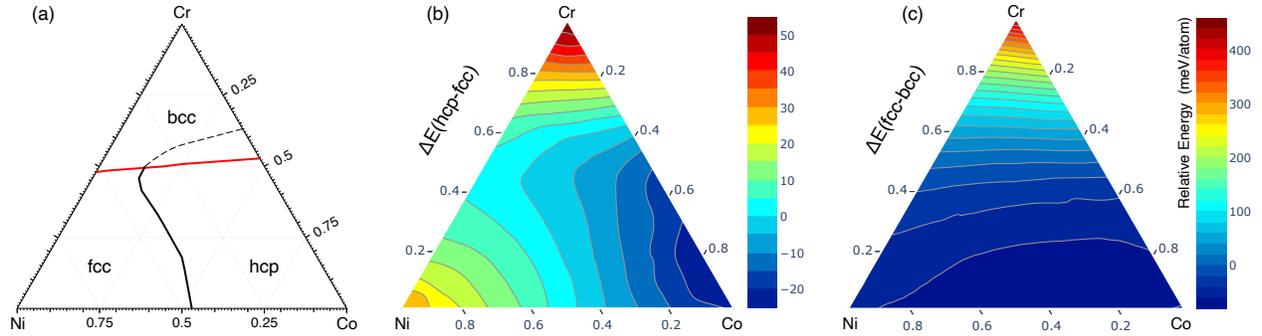

Figure 5. Computed lowest-energy phase and relative energies in Ni-Co-Cr ternary alloys.

On a ternary compositional grid, EoS was computed at each point and the values of equilibrium energy $E_0$ were interpolated using triangulation in Fig. 4. Negative formation energies are present along the Ni-Co binary solid solutions in A1 and A3 phases, see Fig. 6(a). However, at most compositions, including the equiatomic NiCoCr, formation energies are positive; they increase with increasing chromium fraction $x$(Cr) in fcc and hcp phases, see Fig. 4. If formation energies are positive, then the lowest-energy phase is not the ground state, while the ground state is phase segregated.

Energy differences are visualized in Fig. 5. The line depicting zero energy difference in Fig. 5(a) is computed with a higher accuracy compared to the lines in Fig. 5(b,c), obtained from a triangulation and cartesian extrapolation on a sparser grid. A2 (bcc) phase is lower in energy above the red line in Fig. 5(a). The dashed black line shows $E(A1)=E(A3)>E(A2)$; the solid black line shows $E(A1)=E(A3)<E(A2)$; two segments of the red line show $E(A2)=E(A1)<E(A3)$ and $E(A2)=E(A3)<E(A1)$. Red and black lines intersect at the point where all 3 phases have equal positive formation energies, $E(A1)=E(A2)=E(A3)$, see Fig. 5(a).

Energies of binary systems in Fig. 6 and cross sections of the Ni-Co-Cr ternary compositional space in Fig. 7 further elucidate details of the energy-composition dependencies. Equilibrium volumes $V_0$ and formation energies $E_0$ are computed along the selected cross sections of the compositional space on a finer grid in Figs. 6 and 7. Cr stabilizes the bcc (A2) phase, while weakly affecting the hcp-fcc energy difference. This allows for alterations of Cr fraction in the NiCoCr-based middle-entropy alloys. Ni stabilizes the fcc (A1) phase, while Co stabilizes the hcp (A3) phase. Among these 3 elements, Cr atom is the largest, while Ni and Co have comparable sizes.

Impact of the additional chemical elements on the energy landscape is considered in Figs. 8 and 9. Additions of group 4, 5, and 6 metals stabilizes the bcc (A2) phase, see Fig. 8; one can compare their effects with adding more Cr at $x$(Cr)>1/3 in Fig. 7(a). Additions of Al or Fe also stabilizes the bcc phase, see Fig. 9; this effect is expected for the bcc Fe, but is surprising for the fcc Al. $(NiCoCr)_{92}Al_6Ta_2$ medium-entropy alloy was claimed to have improved yield strength, ultimate strength, and tensile ductility at cryogenic temperatures.[45] Addition of both Al and Ti in $(NiCoCr)_{94}Al_3Ti_3$ alloy resulted in formation of $L1_2$ precipitates.[46] At small Al concentrations, Al weakly affect the hcp-fcc energy difference, see Fig. 9(a). In contrast, Cu reduces energy of the fcc phase relative to the hcp even at small concentrations $x$(Cu), see Fig. 9(b). Cu increases formation energy of each phase at small $x$(Cu); $E(x)$ dependence is concave at any $x$ for each phase. From experiment, Cu is known to deteriorate the corrosion resistance and reduce the high-temperature oxidation resistance in $FeCoCrNiCu_x$ alloy.[47]

Iron is magnetic. Fe atoms retain a large magnetic moment M(Fe)≥2.2 Bohr magnetons ($\mu_B$) in the bcc phase at all concentrations $x$(Fe). In contrast, in the close-packed fcc and hcp phases Fe atoms have a magnetic moment at small $x$ but become non-magnetic (with zero atomic magnetic moment) and larger $x$(Fe). A magnetic phase has a larger equilibrium volume $V_0$ and a lower density. Due to this change of the magnetic state, the $V(x)$ curves are not smooth in the close-packed fcc and hcp phases in Fig. 9(c). In a Fe-doped NiCoCr with Fe solute, Fe atoms are magnetic impurities. Magnetic point defects are chemically active; they reduce corrosion resistance of an alloy but can be beneficial for designing catalysts. CrFeCoNi medium-entropy alloy was considered for additive manufacturing.[48]

Chemical homogeneity of an alloy depends on the curvature of energy versus composition dependence $E(x)$.[49] Stabilization of defects by chemistry reduces mobility of those defects and improves creep. Defects can be point (e.g., vacancies), linear (dislocations), planar (stacking faults, twins, grain boundaries, etc.), or bulk (precipitates). Segregation tendency is manifested by the concave $E(x)$ dependence, which makes it energetically favorable to form chemical inhomogeneities. Segregation from bulk results in accumulation of solute near defects that attract solute atoms, acting as a sink. A similar phenomenon was observed in $L1_2$-type precipitates in Ni superalloys.[22]

A concave $E(x)$ dependence is predicted for Zr, Hf, Nb, Ta, Mo, W, and Re, see Fig. 8. In modest amounts, these elements are expected to be beneficial for improving creep in NiCoCr-based alloys.

There are attempts to add simultaneously several elements to NiCoCr. For example, $Ni_{24.27}Co_{35.37}Cr_{20.51}Fe_{9.19}Mo_{4.80}Ti_{4.03}Al_{0.76}Nb_{1.07}$ MP159 alloy was found to have a single-phase fcc structure with a relatively low stacking fault energy; its strength and plasticity were increasing during temperature lowering from 298 K to 77 K.[50]

Addition of large atoms (e.g., La, Ce, and rare earth) can improve strength of alloys, if atomic size mismatch reduces mobility of certain defects. Diffusion of these large and heavy atoms is low, and they can pin the moving defects. Minor inclusion of lanthanum into the high-entropy CrMnFeNi alloy affected corrosion resistance.[51]

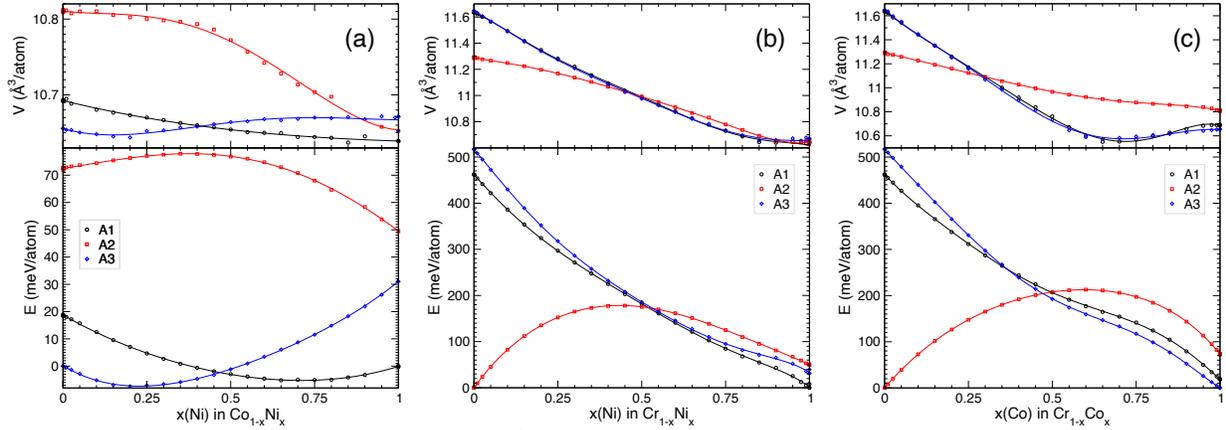

Figure 6. Computed equilibrium volume [Å³/atom] and formation energy [meV/atom] in 3 binary alloys. Formation energies of fcc (A1) Ni$_{0.75}$Co$_{0.25}$ and hcp (A3) Ni$_{0.25}$Co$_{0.75}$ are negative. In bcc (A2) phase, formation energy of elemental Cr is zero, all others are positive.

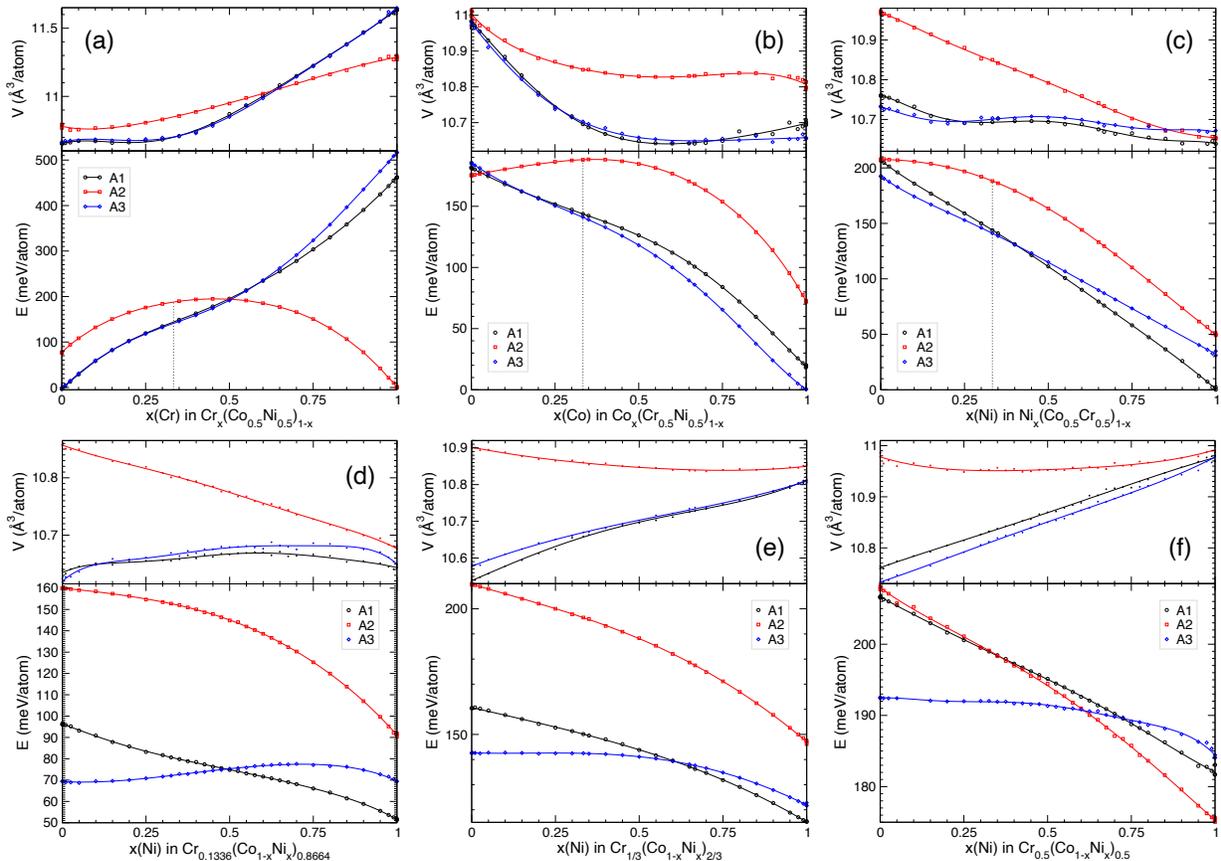

Figure 7. Computed equilibrium volume [Å³/atom] and formation energy [meV/atom] along the median compositions of ternary Ni-Co-Cr alloy (upper) and 3 cross-sections at constant Cr fractions: 0.1336, 1/3, and 1/2 (lower). Energy dependence $E(x)$ in bcc (A2) phase is concave. Energies of the fcc (A1) and hcp (A3) phases intersect.

Interstitial dopants (e.g., boron or carbon) are mobile and interact with defects. Such interactions are beneficial in carbon steels.[52] Boron was claimed to enhance resistance to hydrogen embrittlement in a CrCoNi-based medium-entropy alloy.[53] Addition of carbon into NiCoCr resulted in simultaneous increase of strength and ductility, and also increased the stacking fault energy.[54]

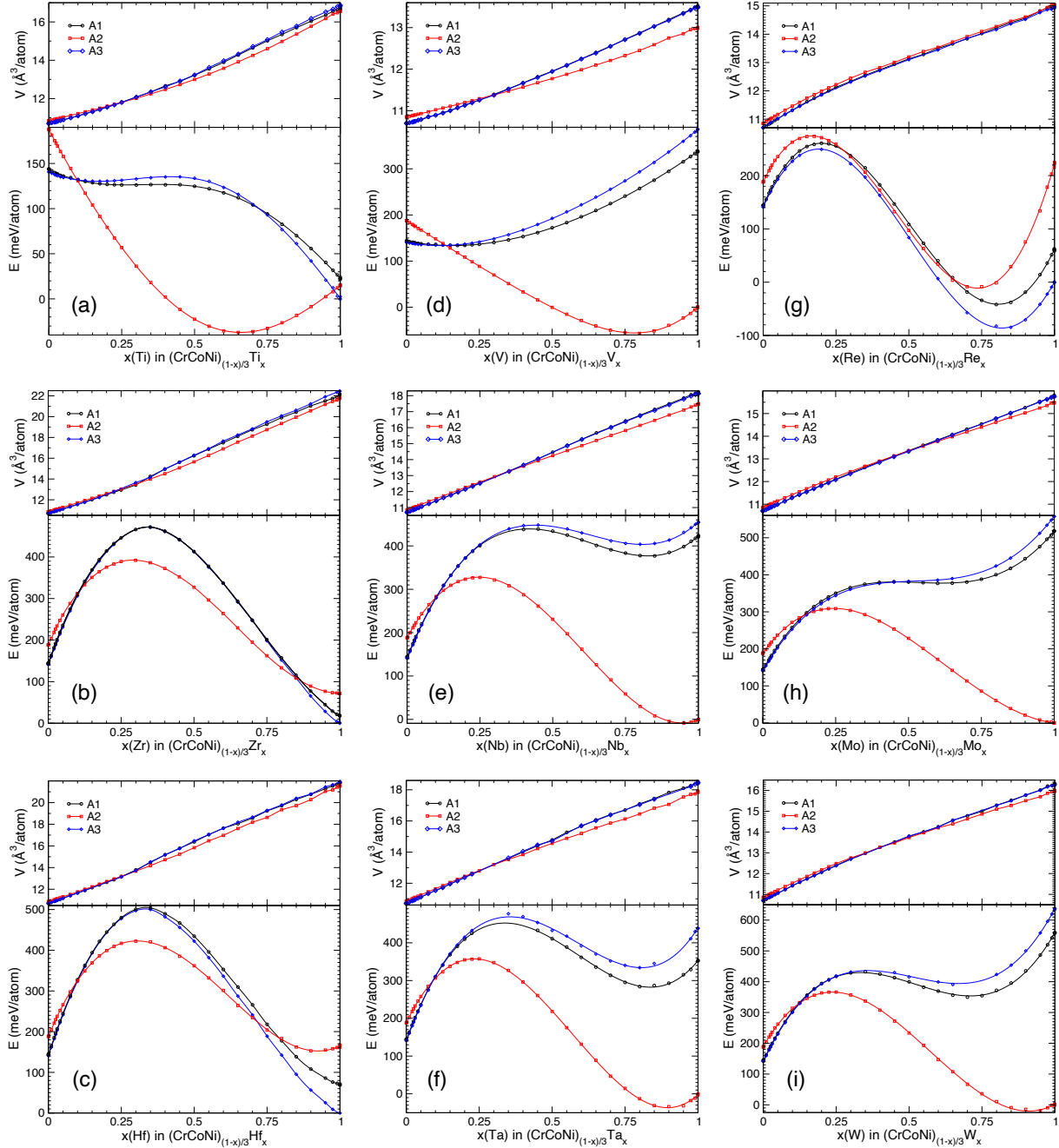

Figure 8. Computed equilibrium volume [Å$^3$/atom] and formation energy [meV/atom] for the selected quaternary alloys (CrCoNi)$_{1-x}$X$_x$ with X=(Ti, Zr, Hf, V, Nb, Ta, Mo, W, Re).

The low stacking fault energy of 18±4 mJ/m$^2$ was measured in NiCoCr sample,[42] which was homogenized at 1473 K for 10 hours, water quenched, cold-rolled, and annealed at 1073 K (800°C) for 1 hour. A similar value of 22±4 mJ/m$^2$ was reported[43] for the NiCoCr sample with the measured composition (at.%) Ni$_{32.85}$Co$_{33.30}$Cr$_{32.53}$Fe$_{0.95}$Mn$_{0.09}$O$_{0.226}$C$_{0.019}$S$_{0.004}$, produced by vacuum induction melting and casting followed by swaging at room temperature and recrystallization at 1173 K for 1 hour.

Fracture toughness of NiCoCr at the cryogenic temperature of 20K was measured to be 459 MPa·m$^{1/2}$ for crack initiation and 540 MPa·m$^{1/2}$ for crack growth after 2.25 millimeters of stable cracking.[55] These exceptionally high values were attributed to a synergy of deformation mechanisms, that included dislocation glide, stacking-fault formation, nano-twinning, and phase transformation. The observed[55] fcc→hcp transformation at cryogenic $T$ is consistent with our calculations, according to which the hcp phase has a slightly lower energy than fcc at 0K in NiCoCr, see Fig. 2 and Table A2. A similar fcc→hcp phase change was reported in several experiments.[4,18] Our results in Figs. 4 and 5 allow to adjust the hcp-fcc energy difference (which is correlated with the stacking fault energy) in a ternary Ni-Co-Cr system, while our data in Figs. 8 and 9 should be useful for considering other chemical elements as dopants.[56]

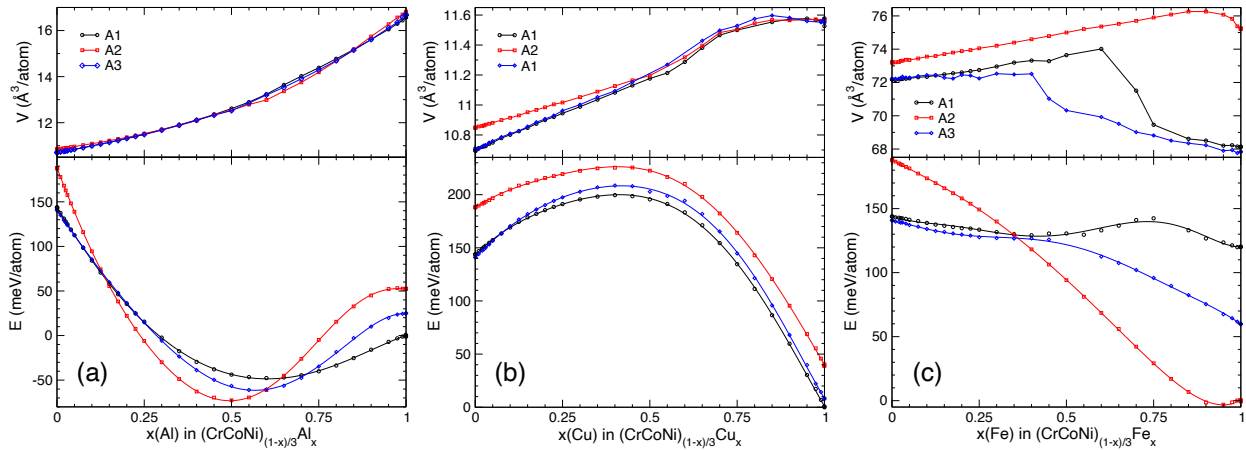

Figure 9. Computed equilibrium volume [Å$^3$/atom] and formation energy [meV/atom] for the quaternary alloys composed by Ni, Co, and Cr with added Al, Cu, or Fe. Convex $E(x)$ points at a mixing tendency for Al (a). Concave $E(x)$ points at a segregation tendency for Cu (b). $V(x)$ dependency is not smooth due to a magnetic phase change in Fe-rich fcc and hcp phases (c).

**4. Summary**:

We combined density functional theory with multiple scattering theory, computed energy-composition dependencies, and investigated the energy landscape in NiCoCr-based middle-entropy alloys. Our results will be useful as a guidance for adjusting composition and designing next-generation alloys with improved strength, creep, ductility, radiation tolerance and corrosion resistance.

**Acknowledgements**:
We acknowledge funding by NASA's Aeronautics Research Mission Directorate (ARMD) via Transformational Tools and Technologies (TTT) Project. We thank Shreyas J. Honrao, Mikhail Mendelev, Valery V. Borovikov, Zhigang Wu, and Anupa R. Bajwa for discussion.

**Appendix**:

The computed lattice constants are compared with experimental ones in Table A1. DFT computations were performed at T=0K, while measurements were done at room temperature (RT). Considering thermal expansion, the agreement is reasonable, as expected for the PBEsol exchange correlation functional. The computed EoS parameters of NiCoCr in 3 phases are compared with those of elemental solids in Table A2. The homogeneously disordered NiCoCr alloy has a positive formation energy (e.g., $E$(hcp) = +141 meV/atom), and the energies of its fcc and hcp phases are very similar: $E$(fcc) – $E$(hcp) < 3 meV/atom.

Table A1. Computed at $T$=0K and experimental[44] (at RT) lattice constants.

| $a$ (Å) | DFT (0K) | Expt. (RT) |
|---|---|---|
| NiCoCr fcc | 3.497 | 3.564 |
| Ni fcc | 3.49 | 3.52 |
| Co hcp | 2.47 | 2.51 |
| Cr bcc | 2.83 | 2.88 |

Table A2. The computed equation of state (EoS) parameters for the equiatomic NiCoCr in fcc (A1), hcp (A3), and bcc (A2) phases: equilibrium formation energy $E_0$ (eV/atom), atomic volume $V_0$ (Å$^3$/atom), bulk modulus $B_0$ (eV/Å$^3$), and pressure derivative $B_0'$ (dimensionless), compared to the computed EoS parameters of the elemental solids (Ni, Co, Cr) in their ground states.

| NiCoCr | A1 (fcc) | A3 (hcp) | A2 (bcc) |
|---|---|---|---|
| $E_0$ (meV/atom) | 143.6 | 141.0 | 188.1 |
| $V_0$ (Å$^3$/atom) | 10.69 | 10.70 | 10.84 |
| $B_0$ (GPa) | 232 | 231 | 228 |
| $B_0'$ | 5.2 | 4.8 | 4.4 |
| | | | |
| Elements | Ni fcc | Co hcp | Cr bcc |
| $E_0$ (meV) | 0 | 0 | 0 |
| $V_0$ (Å$^3$) | 10.64 | 10.66 | 11.29 |
| $B_0$ (GPa) | 218 | 227 | 260 |
| $B_0'$ | 4.7 | 4.5 | 3.7 |

**References**:
1  Senkov, O. N., Miller, J. D., Miracle, D. B. & Woodward, C. Accelerated exploration of multi-principal element alloys with solid solution phases. *Nature Communications* **6**, 6529, doi:10.1038/ncomms7529 (2015).
2  Gludovatz, B. *et al.* Exceptional damage-tolerance of a medium-entropy alloy CrCoNi at cryogenic temperatures. *Nature Communications* **7**, 10602, doi:10.1038/ncomms10602 (2016).
3  Slone, C. E., Miao, J., George, E. P. & Mills, M. J. Achieving ultra-high strength and ductility in equiatomic CrCoNi with partially recrystallized microstructures. *Acta Materialia* **165**, 496-507, doi:10.1016/j.actamat.2018.12.015 (2019).
4  Miao, J. *et al.* The evolution of the deformation substructure in a Ni-Co-Cr equiatomic solid solution alloy. *Acta Materialia* **132**, 35-48, doi:10.1016/j.actamat.2017.04.033 (2017).
5  Miracle, D. B. & Senkov, O. N. A critical review of high entropy alloys and related concepts. *Acta Materialia* **122**, 448-511, doi:10.1016/j.actamat.2016.08.081 (2017).
6  Smith, T. M., Thompson, A. C., Gabb, T. P., Bowman, C. L. & Kantzos, C. A. Efficient production of a high-performance dispersion strengthened, multi-principal element alloy. *Sci Rep-Uk* **10**, 9663, doi:10.1038/s41598-020-66436-5 (2020).

7   Modupeola, D., Patricia, P., Samson, A. & Ntombi, M. in *Aerodynamics* (eds Gorji-Bandpy Mofid & Aly Aly-Mousaad) Ch. 7 (IntechOpen, 2019).
8   Maulik, O., Kumar, D., Kumar, S., Dewangan, S. K. & Kumar, V. Structure and properties of lightweight high entropy alloys: a brief review. *Mater Res Express* **5**, 052001, doi:10.1088/2053-1591/aabbca (2018).
9   Lu, C. *et al.* Irradiation effects of medium-entropy alloy NiCoCr with and without pre-indentation. *Journal of Nuclear Materials* **524**, 60-66, doi:10.1016/j.jnucmat.2019.06.020 (2019).
10  Lu, C. *et al.* High radiation tolerance of an ultrastrong nanostructured NiCoCr alloy with stable dispersed nanooxides and fine grain structure. *Journal of Nuclear Materials* **557**, 153316, doi:10.1016/j.jnucmat.2021.153316 (2021).
11  Zarkevich, N. A. Electricity without Fuel. *Journal of Energy and Power Technology* **2**, 14, doi:10.21926/jept.2001001 (2020).
12  Brif, Y., Thomas, M. & Todd, I. The use of high-entropy alloys in additive manufacturing. *Scripta Mater* **99**, 93-96, doi:10.1016/j.scriptamat.2014.11.037 (2015).
13  Chen, S., Tong, Y. & Liaw, P. K. Additive Manufacturing of High-Entropy Alloys: A Review. *Entropy* **20**, 937 (2018).
14  Ostovari Moghaddam, A., Shaburova, N. A., Samodurova, M. N., Abdollahzadeh, A. & Trofimov, E. A. Additive manufacturing of high entropy alloys: A practical review. *Journal of Materials Science & Technology* **77**, 131-162, doi:10.1016/j.jmst.2020.11.029 (2021).
15  Zhang, W., Chabok, A., Kooi, B. J. & Pei, Y. Additive manufactured high entropy alloys: A review of the microstructure and properties. *Mater Design* **220**, 110875, doi:10.1016/j.matdes.2022.110875 (2022).
16  Argon, A. *Strengthening Mechanisms in Crystal Plasticity*. (Oxford University Press, 2007).
17  Li, M., Guo, Y., Li, W., Zhang, Y. & Chang, Y. Property enhancement of CoCrNi medium-entropy alloy by introducing nano-scale features. *Materials Science and Engineering: A* **817**, 141368, doi:10.1016/j.msea.2021.141368 (2021).
18  Schuh, B. *et al.* Influence of Annealing on Microstructure and Mechanical Properties of a Nanocrystalline CrCoNi Medium-Entropy Alloy. *Materials* **11**, 662 (2018).
19  Smith, T. M. & et al. Microstructural and Mechanical Characterization of a Dispersion Strengthened Medium Entropy Alloy Produced Using Selective Laser Melting. USA patent (2019).
20  Smith, T. M. *et al.* Atomic-scale characterization and modeling of 60° dislocations in a high-entropy alloy. *Acta Materialia* **110**, 352-363, doi:10.1016/j.actamat.2016.03.045 (2016).
21  Zarkevich, N. A., Smith, T. M., Baum, E. N. & Lawson, J. W. Compositional Glass: A State with Inherent Chemical Disorder, Exemplified by Ti-rich Ni3(Al,Ti)1 D024 Phase. *Crystals* **12**, 1049 (2022).
22  Smith, T. M. *et al.* Utilizing local phase transformation strengthening for nickel-base superalloys. *Communications Materials* **2**, 106, doi:10.1038/s43246-021-00210-6 (2021).
23  Zarkevich, N. A. & Johnson, D. D. Between Harmonic Crystal and Glass: Solids with Dimpled Potential-Energy Surfaces Having Multiple Local Energy Minima. *Crystals* **12**, 84 (2022).

24  Jin, K., Gao, Y. F. & Bei, H. Intrinsic properties and strengthening mechanism of monocrystalline Ni-containing ternary concentrated solid solutions. *Materials Science and Engineering: A* **695**, 74-79, doi:10.1016/j.msea.2017.04.003 (2017).
25  Yin, B., Yoshida, S., Tsuji, N. & Curtin, W. A. Yield strength and misfit volumes of NiCoCr and implications for short-range-order. *Nature Communications* **11**, 2507, doi:10.1038/s41467-020-16083-1 (2020).
26  Zarkevich, N. A. Theoretical and computational methods for accelerated materials discovery. *Modern Physics Letters B* **35**, 2130003, doi:10.1142/S0217984921300039 (2021).
27  Zarkevich, N. A. Structural database for reducing cost in materials design and complexity of multiscale computations. *Complexity* **11**, 36-42, doi:10.1002/cplx.20117 (2006).
28  Faulkner, J. S., Stocks, G. M. & Wang, Y. in *Electronic structure of solids*   (IOP Publishing, 2018).
29  Johnson, D. D., Nicholson, D. M., Pinski, F. J., Gyorffy, B. L. & Stocks, G. M. Density-Functional Theory for Random Alloys: Total Energy within the Coherent-Potential Approximation. *Physical Review Letters* **56**, 2088-2091, doi:10.1103/PhysRevLett.56.2088 (1986).
30  Korringa, J. On the calculation of the energy of a Bloch wave in a metal. *Physica* **13**, 392-400, doi:10.1016/0031-8914(47)90013-X (1947).
31  Kohn, W. & Rostoker, N. Solution of the Schr\"odinger Equation in Periodic Lattices with an Application to Metallic Lithium. *Physical Review* **94**, 1111-1120, doi:10.1103/PhysRev.94.1111 (1954).
32  Johnson, D. D., Smirnov, A. V. & Khan, S. N. MECCA: Multiple-scattering electronic-structure calculations for complex alloys. KKR-CPA program. (Iowa State University and Ames Laboratory, Ames, Iowa, 2015).
33  Birch, F. Finite Elastic Strain of Cubic Crystals. *Physical Review* **71**, 809-824, doi:10.1103/PhysRev.71.809 (1947).
34  Murnaghan, F. D. The Compressibility of Media under Extreme Pressures. *Proceedings of the National Academy of Sciences* **30**, 244-247, doi:10.1073/pnas.30.9.244 (1944).
35  Kresse, G. & Hafner, J. Ab initio molecular dynamics for liquid metals. *Phys Rev B* **47**, 558-561, doi:10.1103/PhysRevB.47.558 (1993).
36  Kresse, G. & Hafner, J. Ab initio molecular-dynamics simulation of the liquid-metal--amorphous-semiconductor transition in germanium. *Phys Rev B* **49**, 14251-14269, doi:10.1103/PhysRevB.49.14251 (1994).
37  Kresse, G. & Joubert, D. From ultrasoft pseudopotentials to the projector augmented-wave method. *Phys Rev B* **59**, 1758-1775, doi:10.1103/PhysRevB.59.1758 (1999).
38  Perdew, J. P. *et al.* Restoring the Density-Gradient Expansion for Exchange in Solids and Surfaces. *Physical Review Letters* **100**, 136406, doi:10.1103/PhysRevLett.100.136406 (2008).
39  Pei, Z., Li, R., Gao, M. C. & Stocks, G. M. Statistics of the NiCoCr medium-entropy alloy: Novel aspects of an old puzzle. *npj Computational Materials* **6**, 122, doi:10.1038/s41524-020-00389-1 (2020).
40  Zhang, F. X. *et al.* Local Structure and Short-Range Order in a NiCoCr Solid Solution Alloy. *Physical Review Letters* **118**, 205501, doi:10.1103/PhysRevLett.118.205501 (2017).

41. Zhao, S., Osetsky, Y., Stocks, G. M. & Zhang, Y. Local-environment dependence of stacking fault energies in concentrated solid-solution alloys. *npj Computational Materials* **5**, 13, doi:10.1038/s41524-019-0150-y (2019).
42. Liu, S. F. *et al.* Stacking fault energy of face-centered-cubic high entropy alloys. *Intermetallics* **93**, 269-273, doi:10.1016/j.intermet.2017.10.004 (2018).
43. Laplanche, G. *et al.* Reasons for the superior mechanical properties of medium-entropy CrCoNi compared to high-entropy CrMnFeCoNi. *Acta Materialia* **128**, 292-303, doi:10.1016/j.actamat.2017.02.036 (2017).
44. Hermann, K. in *Crystallography and Surface Structure*    265-266 (Wiley, 2011).
45. Zhang, D. D., Zhang, J. Y., Kuang, J., Liu, G. & Sun, J. Superior strength-ductility synergy and strain hardenability of Al/Ta co-doped NiCoCr twinned medium entropy alloy for cryogenic applications. *Acta Materialia* **220**, 117288, doi:10.1016/j.actamat.2021.117288 (2021).
46. Peng, H. *et al.* Optimization of the microstructure and mechanical properties of electron beam welded high-strength medium-entropy alloy (NiCoCr)94Al3Ti3. *Intermetallics* **141**, 107439, doi:10.1016/j.intermet.2021.107439 (2022).
47. Cai, Y., Chen, Y., Luo, Z., Gao, F. & Li, L. Manufacturing of FeCoCrNiCux medium-entropy alloy coating using laser cladding technology. *Mater Design* **133**, 91-108, doi:10.1016/j.matdes.2017.07.045 (2017).
48. Kuzminova, Y. *et al.* The effect of the parameters of the powder bed fusion process on the microstructure and mechanical properties of CrFeCoNi medium-entropy alloys. *Intermetallics* **116**, 106651, doi:10.1016/j.intermet.2019.106651 (2020).
49. Yibole, H. *et al.* Manipulating the stability of crystallographic and magnetic sub-lattices: A first-order magnetoelastic transformation in transition metal based Laves phase. *Acta Materialia* **154**, 365-374, doi:10.1016/j.actamat.2018.05.048 (2018).
50. Pei, B. *et al.* Excellent combination of strength and ductility in CoNiCr-based MP159 alloys at cryogenic temperature. *Journal of Alloys and Compounds* **907**, 164144, doi:10.1016/j.jallcom.2022.164144 (2022).
51. Sun, Y. P., Wang, Z., Yang, H. J., Lan, A. D. & Qiao, J. W. Effects of the element La on the corrosion properties of CrMnFeNi high entropy alloys. *Journal of Alloys and Compounds* **842**, 155825, doi:10.1016/j.jallcom.2020.155825 (2020).
52. Dwivedi, D., Lepková, K. & Becker, T. Carbon steel corrosion: a review of key surface properties and characterization methods. *Rsc Adv* **7**, 4580-4610, doi:10.1039/C6RA25094G (2017).
53. Chen, X. H. *et al.* Enhanced resistance to hydrogen embrittlement in a CrCoNi-based medium-entropy alloy via grain-boundary decoration of boron. *Materials Research Letters* **10**, 278-286, doi:10.1080/21663831.2022.2033865 (2022).
54. Shang, Y. Y. *et al.* Solving the strength-ductility tradeoff in the medium-entropy NiCoCr alloy via interstitial strengthening of carbon. *Intermetallics* **106**, 77-87, doi:10.1016/j.intermet.2018.12.009 (2019).
55. Liu, D. *et al.* Exceptional fracture toughness of CrCoNi-based medium- and high-entropy alloys at 20 kelvin. *Science* **378**, 978-983, doi:doi:10.1126/science.abp8070 (2022).
56. T.M. Smith *et al.* A 3D Printable Alloy Designed for Extreme Environments. *Nature* (2023).